\documentclass[11pt,a4paper]{article}
\usepackage{amsmath,latexsym,fullpage,amssymb,color}
\usepackage{epic,eepic,ifthen,graphics,epsfig}
\usepackage[english]{babel}
\usepackage{times}
\usepackage{float}
\usepackage{xspace}
\usepackage[T1]{fontenc}
\usepackage{tikz}

\newlength {\squarewidth}

\newtheorem{theorem}{Theorem}
\newtheorem{lemma}{Lemma}

\newcommand{\toto}{xxx}
\newenvironment{proofT}{\noindent{\bf Proof }}
{\hspace*{\fill}$\Box_{Theorem~\ref{\toto}}$\par\vspace{3mm}}
\newenvironment{proofL}{\noindent{\bf Proof }}
{\hspace*{\fill}$\Box_{Lemma~\ref{\toto}}$\par\vspace{3mm}}

\newcounter{linecounter}
\newcommand{\linenumbering}{\ifthenelse{\value{linecounter}<10}
{(\arabic{linecounter})}{(\arabic{linecounter})}}
\renewcommand{\line}[1]{\refstepcounter{linecounter}\label{#1}\linenumbering}
\newcommand{\resetline}[1]{\setcounter{linecounter}{0}#1}
\renewcommand{\thelinecounter}{\ifnum \value{linecounter} > 
9 \else \fi\arabic{linecounter}}
\newfloat{algorithm}{thp}{lop}
\floatname{algorithm}{Algorithm}
\newfloat{construction}{thp}{lop}
\floatname{construction}{Construction}
\newcommand{\Xomit}[1]{}

\newcommand{\KBO}{$k$-BO}
\newcommand{\kbobroadcast}{{\sf kbo\_broadcast}}  
\newcommand{\kbodeliver}{{\sf kbo\_deliver}}
\newcommand{\kscdbroadcast}{{\sf kscd\_broadcast}}  
\newcommand{\kscddeliver}{{\sf kscd\_deliver}}
\newcommand{\scdbroadcast}{{\sf scd\_broadcast}}  
\newcommand{\scddeliver}{{\sf scd\_deliver}}

\newcommand{\KSA}{$k$-SA}

\newcommand{\KSCD}{$k$-SCD}
\newcommand{\KSET}{{\mathit{KSET}}} 
\newcommand{\SNAP}{{\mathit{SNAP}}} 
\newcommand{\ktwospropose}{{\sf k2s\_propose}}
\newcommand{\KSS}{{\mathit{KSS}}} 
\newcommand{\MS}{{\mathit{MS}}} 
\newcommand{\ms}{{\mathit{ms}}}
\newcommand{\fm}{{\mathit{fm}}}
\newcommand{\lm}{{\mathit{lm}}}
\newcommand{\lms}{{\mathit{lms}}}
\newcommand{\fms}{{\mathit{fms}}}
\newcommand{\ums}{{\mathit{ums}}}

\newcommand{\CAMP}{{\cal{CAMP}}}
\newcommand{\RKSA}{{\mathit{RKSA}}} 

\newcommand{\SCD}{{SCD}}

\newcommand{\REG}{\mathit{REG}}

\newcommand{\snapshot}{\sf snapshot}

\newcommand{\propose}{{\sf propose}}

\newcommand{\wait}{{\sf{wait}}}
\newcommand{\return}{{\sf return}} 
\newcommand{\MEM}{\mathit{MEM}}
\newcommand{\mem}{\mathit{mem}}

\newcommand{\todeliver}{\mathit{to\_deliver}} 
 
\newcommand{\delivered}{\mathit{delivered}}

\newcommand{\first}{\mathit{first}}
\newcommand{\rest}{\mathit{rest}}
\newcommand{\prefi}{\mathit{pref_i}}

\newcommand{\width}{{\sf width}}

\newcommand{\decisions}{\mathit{decisions}}

\newcommand{\iinsert}{{\sf insert}}
\newcommand{\wwrite}{{\sf write}}
\newcommand{\mmin}{{\sf min}}
\newcommand{\head}{{\sf head}}
\newcommand{\tail}{{\sf tail}}

\begin{document}

\title{\bf  Which                             Broadcast Abstraction  Captures          $k$-Set Agreement?                  }

  \author{Damien Imbs$^{\circ}$, 
         Achour Most\'efaoui$^{\dag}$,~
         Matthieu Perrin$^{\diamond}$,~
         Michel Raynal$^{\star,\ddag}$\\~\\
   $^{\circ}$LIF, Universit\'e Aix-Marseille, 13288  Marseille, France \\
$^{\dag}$LINA, Universit\'e de Nantes, 44322 Nantes, France \\
$^{\diamond}$IMDEA Software Institute, 28223 Pozuelo de Alarc\'on, Madrid, Spain\\
$^{\star}$Institut Universitaire de France\\
$^{\ddag}$IRISA, Universit\'e de Rennes, 35042 Rennes, France \\
}

\maketitle

\begin{abstract}
It is well-known that consensus (one-set agreement) and total order broadcast 
are equivalent in asynchronous systems prone to process crash failures.
Considering wait-free systems,  this article addresses and answers the following  question: which is the communication abstraction that ``captures'' $k$-set agreement?
To this end, it introduces a new broadcast communication abstraction,
called $k$-BO-Broadcast, which restricts the disagreement on the
local deliveries of the messages that have been broadcast
($1$-BO-Broadcast boils down to total order broadcast).
Hence, in this context, $k=1$ is not a special number,
but only the first integer in  an increasing integer sequence. 

This establishes a new ``correspondence'' between distributed agreement
problems and communication abstractions, which enriches our understanding
of the relations linking fundamental issues of fault-tolerant distributed
computing.  

~\\~\\{\bf Keywords}:
Agreement problem, Antichain, Asynchronous system,
Communication abstraction,  Consensus,
Message-passing system, Partially ordered set,
Process crash, Read/write object, $k$-Set agreement,
Snapshot object,  Wait-free model, Total order broadcast.
\end{abstract}

\section{Introduction}

\paragraph{Agreement problems vs communication abstractions}
Agreement objects are fundamental in the mastering and understanding
of fault-tolerant crash-prone asynchronous distributed systems. The
most famous of them is the {\it consensus} object.  This object
provides processes with a single operation, denoted $\propose()$,
which allows each process to propose a value and decide on (obtain)
a value.  The properties defining this object are the following: If a
process invokes $\propose()$ and does not crash, it decides a value
(termination); No two processes decide different values (agreement);
The decided value was proposed by a process (validity).  This object
has been generalized by S. Chaudhuri in~\cite{C93}, under the name
{\it k-set agreement} ($k$-SA), by weakening the agreement property: the
processes are allowed to collectively decide up to $k$ different
values, i.e., $k$ is the upper bound on the
disagreement allowed on the number of different values that can be decided.
The smallest value $k=1$ corresponds to consensus.

On another side, communication abstractions allow processes
to exchange data and coordinate, according to some message
communication patterns. Numerous communication abstractions
have been proposed. Causal message delivery~\cite{BJ87,RST91},
total order broadcast, FIFO broadcast, to cite a  few (see the
textbooks~\cite{AW04,L96,R10,R13}).
In a very interesting way, it appears that some high level 
communication abstractions  ``capture'' exactly the essence of some
agreement objects, see  Table~\ref{table1}.
The most famous --known since a long time-- 
is the {\it Total Order broadcast} abstraction
which, on one side,  allows an easy implementation of a consensus object,
and, on an other side, can be implemented from consensus objects.
A more recent example is the SCD-Broadcast abstraction that we introduced
in~\cite{IMPR17} (SCD stands for {\it Set Constrained Delivery}).
This communication abstraction allows a very easy implementation of an
atomic (Single Writer/Multi Reader or Multi Writer/Multi Reader)
snapshot object (as defined in~\cite{AADGMS93}), and can also be
implemented from snapshot objects. Hence, as shown in~\cite{IMPR17},
SCD-Broadcast and snapshot objects are the two sides of a same
``coin'': one side is concurrent object-oriented, the other side is
communication-oriented, and none of them is more computationally powerful
than the other in asynchronous wait-free systems
(where ``wait-free'' means ``prone to any number of process crashes'').

\begin{table}[ht]
\begin{center}
\renewcommand{\baselinestretch}{1}
\small
\begin{tabular}{|c|c|}
\hline
Concurrent object &   Communication abstraction \\
\hline
\hline
Consensus~ &   Total order broadcast~\cite{CT96}\\
\hline
Snapshot object~\cite{AADGMS93,A94} (and R/W register)
                                    &   \SCD-broadcast~\cite{IMPR17}\\
\hline
$k$-set agreement object  $(1\leq k <n)$ &   $k$-BO-broadcast~(this paper)\\
\hline 
\end{tabular}
\end{center}
\vspace{-0.3cm}
\caption{Associating agreement objects and communication abstractions}
\label{table1} 
\end{table}

\paragraph{Aim and content of the paper}
As stressed in~\cite{FM03}, Informatics is a science of abstractions.
Hence, this paper continues our quest relating communication abstractions
and agreement objects. It focuses on $k$-set agreement in asynchronous 
wait-free systems. More precisely, the paper introduces the $k$-BO-broadcast
abstraction and shows that it matches $k$-set agreement in these systems. 

$k$-BO-broadcast is a {\it Reliable Broadcast} communication
abstraction~\cite{AW04,L96,R10,R13}, enriched with an additional
property which restricts the disagreement on message receptions among
the processes.  Formally, this property is stated as a constraint on
the width of a partial order whose vertices are the messages, and
directed edges are defined by local message reception orders. 
This width is upper bounded by $k$.  For the extreme case
$k=1$, $k$-BO-broadcast boils down to total order broadcast.

\begin{figure}[htb]
\centering{
\scalebox{0.3}{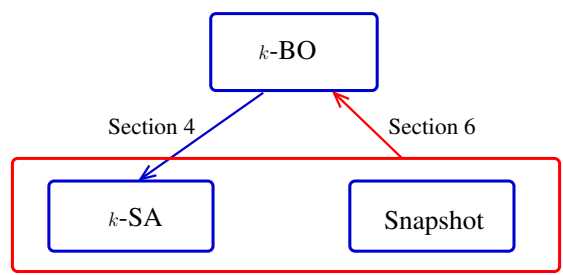}
\caption{Global picture}
\label{fig:global-picture-1}
}
\end{figure}

The correspondence linking $k$-BO-broadcast and $k$-set agreement,
established in the paper, is depicted in
Figure~\ref{fig:global-picture-1}.  The algorithm building $k$-SA on
top of the $k$-BO-broadcast is surprisingly simple (which was our
aim\footnote{This is important, as communication abstractions
  constitute the basic programming layer on top of which distributed
  applications are built.}).  In the other direction, we show that
$k$-BO-broadcast can be implemented in wait-free systems enriched with
$k$-SA objects and snapshot objects. (Let us recall that snapshot
objects do not require additional computability power to be built on
top of wait-free read/write systems.)  This direction is not as simple
as the previous one.  It uses an intermediary broadcast communication
abstraction, named $k$-SCD-broadcast, which is a natural and simple
generalization of the SCD-broadcast introduced in~\cite{IMPR17}.

\paragraph{Roadmap}
The paper is composed of~\ref{sec:conclusion} sections.
Section~\ref{sec:base-models} presents the basic crash-prone
process model, the snapshot object, and $k$-set agreement.
Section~\ref{sec:kBO} defines the $k$-BO broadcast
abstraction and presents a characterization of it.  Then,
Section~\ref{sec:kBO-to-kSA} presents a simple algorithm implementing
$k$-set agreement on top of the $k$-BO broadcast abstraction.
Section~\ref{sec:SCD-to-kBO} presents another simple algorithm
implementing $k$-BO broadcast on top of the $k$-SCD-broadcast
abstraction.  Section~\ref{sec:kSA-snapshot-to-kSCD} presents two
algorithms whose combination implements $k$-SCD-broadcast on top of
$k$-set agreement and snapshot objects. Finally,
Section~\ref{sec:conclusion} concludes the paper.  A global view on the way
these constructions are related is presented in  Figure 2 of the conclusion.

\section{Process Model, Snapshot, and  and $k$-Set Agreement}
\label{sec:base-models} 
\paragraph{Process and failure model} 
The computing model is composed of a set of $n$ asynchronous
sequential processes, denoted $p_1$, ..., $p_n$. ``Asynchronous'' means
that each process proceeds at its own speed, which can be arbitrary
and always remains  unknown to the other processes.  

A process may halt prematurely (crash failure), but it executes 
its local algorithm correctly until its possible crash. 
It is assumed that up to $(n-1)$ processes may crash in a run 
(wait-free failure model).
A process that crashes in a run is said to be {\it faulty}. Otherwise, 
it is {\it non-faulty}. Hence a faulty process behaves as a non-faulty process
until it crashes. 

\paragraph{Snapshot object}
The snapshot object was introduced in~\cite{AADGMS93,A94}.  A snapshot
object is an array $\REG[1..n]$ of single-writer/multi-reader
atomic read/write registers which
provides the processes with two operations, denoted $\wwrite()$ and
$\snapshot()$. Initially $\REG[1..n]=[\bot,\ldots,\bot]$. The invocation
of $\wwrite(v)$ by a process $p_i$ assigns $v$ to $\REG[i]$, and the
invocation of $\snapshot()$ by a process $p_i$ returns the value of
the full array as if the operation had been executed instantaneously.
Said another way, the operations $\wwrite()$ and $\snapshot()$ are
atomic, i.e., in any execution of a snapshot object,
its operations $\wwrite()$ and $\snapshot()$ are linearizable.

If there is no restriction on the number of invocations of $\wwrite()$
and $\snapshot()$ by each process, the snapshot object is multi-shot.
Differently, a one-shot snapshot object is such that each process
invokes once each operation, first $\wwrite()$ and then
$\snapshot()$. The one-shot snapshot objects satisfy a very nice and
important property, called {\it Containment}.  Let $reg_i[1..n]$ be
the vector obtained by $p_i$, and $view_i= \{\langle
reg_i[x],i\rangle~|~reg_i[x] \neq\bot\}$.  For any pair of processes
$p_i$ and $p_j$ which obtain $view_i$ and $view_j$ respectively, we
have $(view_i\subseteq view_j)\vee(view_j\subseteq view_i)$.

Implementations of  snapshot objects on top of
read/write atomic registers have been proposed
(e.g.,~\cite{AADGMS93,A94,IR12,ICMT94}). 
The ``hardness'' to build snapshot objects in read/write systems and
associated lower bounds are presented in the  survey~\cite{E05}.

\Xomit{\paragraph{Message-passing communication model}
In this communication model, each pair of processes communicate by
sending and receiving messages through two uni-directional channels,
one in each direction. Hence, the communication network is a complete
network: any process $p_i$ can directly send a message to any process
$p_j$ (including itself).  A process $p_i$ invokes the operation
``${\sf send}$ {\sc type}($m$) ${\sf to}$ $p_j$'' to send to $p_j$ the
message $m$, whose type is {\sc type}.  The operation ``${\sf
  receive}$ {\sc type}() ${\sf from}$ $p_j$'' allows $p_i$ to receive
from $p_j$ a message whose type is {\sc type}.

Each channel is reliable (no loss, corruption, nor creation of messages),
not necessarily first-in/first-out, and asynchronous (while the transit time 
of each message is finite, there is no upper bound on message transit times).
Let us notice that, due to process and message asynchrony, no process can
know if another process crashed or is only very slow.
}
\label{sec:kSA}

\paragraph{$k$-Set agreement}
$k$-Set agreement (\KSA) was introduced by S. Chaudhuri in~\cite{C93}
(see~\cite{R16} for a survey of $k$-set agreement in various contexts). 
Her aim was to investigate the impact of the maximal
number of process failures ($t$) on the agreement degree ($k$)
allowed to the processes, where the smaller the value of $k$,
the stronger the agreement degree. The maximal agreement degree
corresponds to  $k=1$ (consensus).

\KSA\ is a one-shot agreement problem, which provides the processes
with a single operation denoted $\propose()$.
When a process $p_i$ invokes  $\propose(v_i)$, we say that
it "proposes  value $v_i$''.  This operation returns a value $v$.
We then say that the invoking process ``decides $v$'', and
``$v$ is a decided value''. Assumed that all non-faulty processes invoke
$\propose()$, \KSA\ is defined by the following properties. 

\begin{itemize}
\vspace{-0.2cm}
\item Validity.
If a process decides a value $v$, $v$ was proposed by a process. 
\vspace{-0.2cm}
\item Agreement.
At most $k$ different values are decided by the processes.
\vspace{-0.2cm}
\item Termination.
Every non-faulty process decides a value.   
\end{itemize}

\Xomit{From a computability point of view \KSA\  ca be implemented in
a read/write system  only if $t<k$ ~\cite{BG93,HS99,SZ00}. 
As $t<n/2$ is a necessary and sufficient condition
to implement a read/write register in ${\CAMP}_{n,t}[\emptyset]$ \cite{ABD95}, 
\KSA\  can be implemented in ${\CAMP}_{n,t}[t<n/2]$ when $t<k$. 
The enrichment of  ${\CAMP}_{n,t}[\emptyset]$ with appropriate failure
detectors to solve \KSA\ is addressed in~\cite{BR11,DFGT08,MRS11,MRS12}. 

In read/write asynchronous systems where up to $t<n$ processes may
crash, the combination of the results presented in~\cite{BRS15,DFKR15}
shows that repeated $k$-set agreement can be
obstruction-free\footnote{ Obstruction-freedom is a weak termination
  property. A non-faulty process is required to terminate if it
  executes solo during a long enough period~\cite{HLM03}.}  solved
with $(n-k+1)$ atomic read/write registers, which is optimal
(using a sequence of $x$ $k$-set agreement instances would require
$x(n-k+1)$  atomic read/write registers). 
}

\paragraph{Repeated $k$-set agreement}
This agreement abstraction is a simple 
generalization of  $k$-set agreement,  which aggregates a 
sequence of  $k$-set agreement instances into a single object. 
Hence given such an object  $\RKSA$, 
a process $p_i$ invokes sequentially 
$\RKSA.{\sf  propose}(sn_i^1,v_i^{1})$, then
$\RKSA.{\sf propose}(sn_i^2,v_i^{2}$, ...,
$\RKSA.{\sf propose}(sn_i^x,v_i^{x})$, etc,
where $sn_i^1,sn_i^2,\ldots, sn_i^x, \ldots$ are
increasing (not necessarily consecutive) sequence numbers,
and $v_i^{x}$ is the  value proposed by $p_i$ to the instance number $sn_i^x$. 
Moreover, the sequences of  sequence numbers used by two processes
are sub-sequences of $0$, $1$, $2$,  etc., , but are not necessarily
the same sub-sequence.

\section{The $k$-BO-Broadcast Abstraction}
\label{sec:kBO}

\paragraph{Communication operations}
The $k$-Bounded Ordered broadcast abstraction (\KBO-Broadcast)
provides the processes with
two operations, denoted $\kbobroadcast()$ and  $\kbodeliver()$.
The first operation takes a message as input  parameter. 
The second one returns a message to the process that invoked it. 
Using a classical terminology, when a process invokes $\kbobroadcast(m)$,
we say that it ``kbo-broadcasts the message $m$''. Similarly, when it invokes 
$\kbodeliver()$ and obtains a  message $m$, we say that
it ``kbo-delivers  $m$''; in the operating system parlance,
$\kbodeliver()$ can be seen as an {\it up call}
(the messages kbo-delivered are deposited in a buffer, which is 
accessed by the application according to its own code).

\paragraph{The partial order  $\mapsto$}
Let us first remember a few graph definitions associated with
partially ordered sets.  An {\it antichain} is a subset of a partially
ordered set such that any two elements in the subset are incomparable,
and a {\it maximum antichain} is an antichain that has the maximal
cardinality among all antichains.  The {\it width} of a partially
ordered set is the cardinality of a maximum antichain.

Let $\mapsto_i$ be the local message delivery order at a process
$p_i$ defined as follows: $m \mapsto_i m'$ if $p_i$ kbo-delivers the
message $m$ before it kbo-delivers the message $m'$.
Let  $\mapsto \stackrel{def}{=} \cap_i \mapsto_i$. 
This relation defines a partially ordered set relation which captures
the order on message kbo-deliveries on which all processes agree.
In the following we use the same notation ( $\mapsto$) for the
relation and the associated partially ordered graph.  Let
$\width(\mapsto)$ denote the width of the partially ordered graph $\mapsto$.

\paragraph{Properties on the operations}
\KBO-broadcast is defined by the following
set of properties, where we assume --without loss of generality--
that all the messages that are kbo-broadcast are different. 
\begin{itemize}
\vspace{-0.1cm}
\item KBO-Validity.
  Any  message  kbo-delivered has been  kbo-broadcast by a process.
\vspace{-0.2cm}
\item KBO-Integrity.
  A message is kbo-delivered at most once by each process.
\vspace{-0.2cm}
\item KBO-Bounded.  $\width(\mapsto)\leq k$.
\vspace{-0.2cm}
\item KBO-Termination-1.
  If a non-faulty process kbo-broadcasts a message $m$,
  it terminates its kbo-broadcast invocation and kbo-delivers $m$.
\vspace{-0.2cm}
\item KBO-Termination-2.
  If a process kbo-delivers a message $m$, every  non-faulty
  process kbo-delivers $m$.
\end{itemize}

The reader can easily check that the Validity, Integrity, 
Termination-1, and Termination-2 properties define
{\it Uniform  Reliable Broadcast}.

The KBO-Bounded property, which gives its meaning to
\KBO-broadcast, is new. Two processes $p_i$ and $p_j$ {\it disagree}
on the kbo-deliveries of the messages $m$ and $m'$ if $p_i$
kbo-delivers $m$ before $m'$, while $p_j$ kbo-delivers $m'$ before
$m$. Hence we have neither $m \mapsto m'$ nor $m'\mapsto m$.

$k$-Bounded Order captures the following constraint: processes can
disagree on message sets of size at most $k$. (Said differently, there
is no message set $ms$ such that $|ms|>k$ and 
for each pair of messages  $m, m' \in ms$, there are two processes
$p_i$ and $p_j$ that disagree on their kbo-delivery order.)
Let us consider the following example to illustrate this constraint.

\paragraph{An example}
Let $m_1$, $m_2$, $m_3$, $m_4$, $m_5$, and $m_6$, be messages that have
been kbo-broadcast by different processes.  Let us consider the following
sequences of kbo-deliveries by the processes $p_1$, $p_2$ and $p_3$.
\begin{itemize}
\vspace{-0.2cm}
\item at $p_1$: $m_1$, $m_2$, $m_3$, $m_4$, $m_5$ $m_6$.
\vspace{-0.2cm}
\item
  at $p_2$: $m_2$, $m_1$, $m_5$, $m_3$, $m_4$, $m_6$. 
\vspace{-0.2cm}
\item
    at $p_3$: $m_2$, $m_3$, $m_1$, $m_5$, $m_4$, $m_6$.  
\end{itemize}

The set of messages $\{m_1,m_2\}$ 
is  such that  processes disagree on their kbo-delivery order.
We have the same for the sets of messages   $\{m_1,m_3\}$ and $\{m_4,m_5\}$.
It is easy to see that, when considering the set $\{m_1,m_2,m_3,m_4\}$,
the message $m_4$ does not create disagreement with respect to
the messages in the  set $\{m_1,m_2,m_3\}$.

The reader can check that there is no  set of cardinality greater than
$k=2$ such that processes disagree on all the pairs of messages they contain. 
On the contrary, when looking at the message sets of size $\leq 2$,
disagreement is allowed, as shown by the sets  of messages $\{m_1,m_2\}$,
$\{m_1,m_3\}$,  and $\{m_4,m_5\}$. In conclusion, these sequences of
kbo-deliveries are compatible with 2-BO broadcast.

Let us observe that if two processes disagree on the kbo-deliveries of
two messages $m$ and $m'$, these messages define an antichain of size $2$. 
It follows that $1$-BO-broadcast is nothing else than total order
broadcast (which is computationally equivalent to Consensus~\cite{CT96}),
while $k=n$ imposes no constraint on message deliveries.

\paragraph{Underlying intuition: the non-deterministic  $k$-TO-channel notion}
Let us define the notion of a  {\it non-deterministic  $k$-{\em TO}-channel}
as follows.  There are $k$ different broadcast channels,
each  ensuring  total order delivery on the messages broadcast through it. 
The invocation of  $\kbobroadcast(m)$ by a process entails a
broadcast on one and only one of these broadcast channels, but
the channel is selected by an underlying daemon, and the issuing process
never knows which channel has been selected for its invocation.

Let us consider the previous example, with $k=2$. Hence,  there are
two TO-channels, $channel[1]$ and  $channel[2]$. As shown by the following figure, 
they contained the following sequences of messages:
$channel[1]~=~ m_1,~m_5$,$~m_6$ and   $channel[2]~=~ m_2,~m_3,~m_4$.
On this figure, encircled grey areas represent maximum antichains.

\begin{center}\vspace{-0.4cm}
\begin{tikzpicture}
 
      \fill[black!10,rounded corners=20] (1+0,0-1) -- (1+1,0-0) -- (0-0,1+1) -- (0-1,1+0) -- cycle;
      \fill[black!10,rounded corners=20] (3-1,0-0) -- (3+0,0-1) -- (4+1,1+0) -- (4+0,1+1) -- cycle;
      \fill[black!10,rounded corners=20] (5+0,0-1) -- (5+1,0-0) -- (4-0,1+1) -- (4-1,1+0) -- cycle;

      \draw[dashed,rounded corners=20] (1+0,0-1) -- (1+1,0-0) -- (0-0,1+1) -- (0-1,1+0) -- cycle;
      \draw[dashed,rounded corners=20] (3-1,0-0) -- (3+0,0-1) -- (4+1,1+0) -- (4+0,1+1) -- cycle;
      \draw[dashed,rounded corners=20] (5+0,0-1) -- (5+1,0-0) -- (4-0,1+1) -- (4-1,1+0) -- cycle;
 
      \draw (-1,1) node[left]{$\mathit{channel}[1]$};
      \draw (-1,0) node[left]{$\mathit{channel}[2]$};
 
      \draw (0,1) node{$\bullet$};
      \draw (0,1) node[above]{$m_1$};
 
      \draw (1,0) node{$\bullet$};
      \draw (1,0) node[below]{$m_2$};
 
      \draw (3,0) node{$\bullet$};
      \draw (3,0) node[below]{$m_3$};
 
      \draw (5,0) node{$\bullet$};
      \draw (5,0) node[below]{$m_4$};
 
      \draw (4,1) node{$\bullet$};
      \draw (4,1) node[above]{$m_5$};
 
      \draw (6,1) node{$\bullet$};
      \draw (6,1) node[above]{$m_6$};
 
      \draw (0.5,0.5) node{\tiny $\begin{array}{c} m_1 \mapsto_1 m_2 \\ m_2 \mapsto_2 m_1 \end{array}$};

      \draw[|->] (1.2,0) -- (2.8,0);
      \draw[|->] (3.2,0) -- (4.8,0);
      \draw[|->] (0.2,1) -- (3.8,1);
      \draw[|->] (4.2,1) -- (5.8,1);
      \draw[|->] (1.2,0.1) -- (3.8,0.9);
      \draw[|->] (5.2,0.2) -- (5.8,0.8);
      \draw[|->] (0.2,0.9) -- (4.8,0.1); 
    \end{tikzpicture}
\end{center}

It is easy to check that the sequence of messages delivered at any process
$p_i$ is a merge of the sequences associated with these two channels.

The assignment of messages to channels is not necessarily unique,
it depends on the behavior of the daemon. Considering $k=3$ and
a third channel $channel[3]$, 
let us observe that the same message kbo-deliveries at $p_1$, $p_2$,
and $p_3$, could have been obtained by the following channel selection by
the daemon: 
$channel[1]$ as before, $channel[2]~=~m3,~m4$, and $channel[3]~=~m2$.
Let us observe that,  with $k=3$ and this daemon last behavior, the
message kbo-delivery
$m_3$, $m_1$, $m_5$, $m_4$, $m_2$, $m_6$  would also be non-faulty at $p_3$.

\paragraph{A characterization}
The previous  non-deterministic  $k$-TO-channel interpretation of \KBO-broadcast is captured by the following characterization theorem.

\begin{theorem}
\label{characterization-theorem}
A non-deterministic $k$-{\em TO-channel}  and the {\em \KBO-broadcast}
communication abstraction have the same computational power.  
\end{theorem}

\begin{proofT}
Direction \KBO-broadcast to $k$-TO-channel.  Let us consider the
partial order $\mapsto$ on message kbo-deliveries.  As
$\width(\mapsto) \leq k$, it follows from Dilworth's Theorem
(\cite{D50}) that there is a partition of $\mapsto$ in at most $k$
different chains of messages. Let us associate a channel with each of
these chains. Due to the definition of $\mapsto~ = \cap_i \mapsto_i$,
all the processes kbo-deliver the messages of a given chain in the
same order. It follows that the channel associated with this chain
implements a total order broadcast, and this is true for all the
channels.

Direction  $k$-TO-channel to \KBO-broadcast. 
Let us consider an antichain $\langle m_1,\cdots, m_\ell \rangle$.
Hence,  any two different messages $m_x$ and $m_y$ of it
are such that we have neither $m_x \mapsto m_y$,  nor $m_y \mapsto m_x$.
This  means that $m_x$ and $m_y$ are kbo-delivered in
different order by at least two processes (disagreement).
As each of the $k$ channels ensures total order delivery, it follows that 
$m_x$ and $m_y$ have been broadcast on different channels, and consequently
we have $\ell\leq k$. 
\renewcommand{\toto}{characterization-theorem}
\end{proofT}

\paragraph{Remark}
It is important to see that \KBO-broadcast and $k$-TO-channels are
not only computability equivalent but are two statements of the very
same communication abstraction (there is no way to distinguish them
from a process execution point of view).

\section{From $k$-BO-Broadcast to  Repeated $k$-Set Agreement}
\label{sec:kBO-to-kSA}

Algorithm~\ref{algo:kbo-to-rksa} 
implements repeated $k$-set agreement in a wait-free system enriched
with $k$-BO-Broadcast.
Its simplicity demonstrates the very {\it high abstraction level}
provided by $k$-BO-Broadcast. All ``implementation details'' 
are hidden inside its implementation (which has to be designed only once,
and not for each use of $k$-BO-Broadcast in different contexts).
In this sense, $k$-BO-Broadcast is the abstraction communication which
captures the essence of (repeated) $k$-set agreement.

\paragraph{Local data structure}
Each process $p_i$ manages a set denoted $\decisions_i$ (initially
empty) which contains at most one pair $\langle nb,-\rangle$ per
sequence number $nb$; $\langle nb,v\rangle \in \decisions_i$ means
that value $v$ can be returned by $p_i$ when it invokes $\propose(nb,-)$.

\paragraph{Process behavior}
Let us assume that a process $p_i$ invokes $\propose(sn_i^1,-)$,
$\propose(sn_i^2,-)$, etc. When it invokes $\propose(nb,v)$, $p_i$
kbo-broadcasts a message
containing the pair $\langle nb,v\rangle$ and waits until a pair
$\langle nb,- \rangle$ appears in its local set $\decisions_i$
(lines~\ref{kbo-kset-01}).  When this occurs, it
returns the value $x$ contained in this pair, which is then suppressed
from the set $\decisions_i$ (lines~\ref{kbo-kset-02}).
When a process $p_i$ kbo-delivers a message $m = \langle sn, x
\rangle$ it inserts in $\decisions_i$ only if no message carrying the
same sequence number $sn$ has previously been inserted in
$\decisions_i$ (lines~\ref{kbo-kset-02}).
Let us observe that this algorithm is purely based on
the $k$-BO-Broadcast communication abstraction.  
 
\begin{algorithm}[h]
\centering{\fbox{
\begin{minipage}[t]{150mm}
\footnotesize 
\renewcommand{\baselinestretch}{2.5}
\resetline
\begin{tabbing}
aaaaaa\=aa\=aaa\=aaaaaa\=\kill

{\bf operation} $\propose(nb,v)$ {\bf is}\\

\line{kbo-kset-01} \>  
   $\kbobroadcast(\langle nb, v \rangle)$;
             $\wait(\exists ~\langle nb, x \rangle \in \decisions_i)$;
       $\return (x)$.~\\~\\

{\bf when a message} $\langle sn, x \rangle$ {\bf is kbo-delivered} {\bf do} \\

\line{kbo-kset-02} \> {\bf if}
   $(\langle sn, - \rangle \mbox{ never added to  } \decisions_i)$ 
         {\bf then} $\decisions_i.\iinsert(\langle sn, x \rangle)$    
         {\bf end if}.

\end{tabbing}
\end{minipage}
}
\caption{From $k$-BO-broadcast to repeated $k$-set agreement}
\label{algo:kbo-to-rksa}
}
\end{algorithm}

\begin{lemma}
\label{lemma:validity-kbo-to-kset}
If the invocation of  $\propose(nb,v)$ returns $x$ to a process, 
some process invoked  $\propose(nb,x)$.
\end{lemma}

\begin{proofL}
Let us assume that the invocation  $\propose(nb,-)$ issued by a process
$p_i$ returns the value $x$. It follows 
that $\langle nb, x \rangle \in \decisions_i$. Consequently, the pair
$\langle nb, x \rangle$ has previously been inserted in  $\decisions_i$
at line~\ref{kbo-kset-02}, when $p_i$ kbo-delivered the message carrying
this pair. By the KBO-Validity property,
this message has previously been  kbo-broadcast by some process $p_j$ at 
at line~\ref{kbo-kset-01}, which concludes the proof of the lemma.
\renewcommand{\toto}{lemma:validity-kbo-to-kset}
\end{proofL}

\begin{lemma}
\label{lemma:termination-kbo-to-kset}
If a non-faulty process invokes  $\propose(nb,-)$, it  eventually decides a
value $x$ such that $\langle nb, x \rangle$ is the first (and only) message
$\langle nb, - \rangle$ it kbo-delivers.
\end{lemma}

\begin{proofL}
  If a non-faulty process $p_i$ invokes $\propose(nb,v)$, it follows from the
  KBO-Termination-1 property that  it eventually kbo-delivers
  the  message carrying $\langle nb, v \rangle$.
  Hence, if the  pair  $\langle nb, v \rangle$ is the first pair
  with sequence number $nb$  kbo-delivered by $p_i$, it follows from the
  predicate of line~\ref{kbo-kset-02} that  $\langle nb, v \rangle$ is
  inserted  in $\decisions_i$. Otherwise, another pair$\langle nb,- \rangle$ 
  was previously inserted in  $\decisions_i$.
  Hence, one (and only one) pair with  sequence number  $nb$ is 
  inserted in  the set $\decisions_i$. The lemma  follows then
  from the waiting predicate of line~\ref{kbo-kset-01}.
\renewcommand{\toto}{lemma:termination-kbo-to-kset}
\end{proofL}

\begin{lemma}
\label{lemma:agreement-kbo-to-kset}
The set of values returned by the invocations of $\propose(nb,-)$
contains at most $k$ different values.
\end{lemma}

\begin{proofL}
  Let $\Pi_{nb}$ be the set of  processes returning a value
  from their invocations  $\propose(nb,-)$.
  For each $p_i \in \Pi_{nb}$, let 
  $\langle nb, x_i \rangle$ denote the first message
  $\langle nb, - \rangle$ received by $p_i$.  By
  Lemma~\ref{lemma:termination-kbo-to-kset},
  $X_{nb}~=~\{x_i : p_i \in \Pi_{nb}\}$ is the set of all values
  returned by the invocations of   $\propose(nb,-)$.

  For any pair $x_i$ and $x_j$ of distinct elements of $X_{nb}$, 
  we have that $p_i$ kbo-delivered $x_i$ before $x_j$, and  
  $p_j$ kbo-delivered $x_j$ before $x_i$. Hence, 
  $\langle nb, x_j \rangle  \not\mapsto_i \langle nb, x_i \rangle$ and 
  $\langle nb, x_i \rangle  \not\mapsto_j \langle nb, x_j \rangle$,
  which means $\langle nb, x_i  \rangle$
  and $\langle nb, x_j \rangle$ are not ordered by $\mapsto$. 
  Therefore, $\{\langle nb, x_i \rangle : p_i \in
  \Pi_{nb}\}$ is an antichain of $\mapsto$.
  It then follows from  the
  KBO-Bounded property that $|\{x_i : p_i \in \Pi_{nb}\}| = |\{\langle nb,
  x_i \rangle : p_i \in \Pi_{nb}\}| \le k$.
  \renewcommand{\toto}{lemma:agreement-kbo-to-kset}  
\end{proofL}

\begin{theorem} 
\label{theorem:proof-kbo-to-kset}
Algorithm~{\em{\ref{algo:kbo-to-rksa}}} implements repeated $k$-set
agreement in any system model enriched with
the communication abstraction $\mbox{\em \KBO-broadcast}$.
\end{theorem}

\begin{proofT}
  The proof follows from
  Lemma~\ref{lemma:validity-kbo-to-kset} (validity),
  Lemma~\ref{lemma:agreement-kbo-to-kset} (agreement), and
  Lemma~\ref{lemma:termination-kbo-to-kset} (termination). 
\renewcommand{\toto}{theorem:proof-kbo-to-kset}
\end{proofT}

\section{From $k$-SCD-Broadcast to $k$-BO-Broadcast}
\label{sec:SCD-to-kBO}

\subsection{The intermediary $k$-SCD-Broadcast abstraction}
This  communication abstraction is a simple strengthening of
the SCD-Broadcast abstraction introduced in~\cite{IMPR17},
where it is shown that SCD-Broadcast and snapshot objects have the same
computability power. SCD stands for Set Constrained Delivery). 

\paragraph{SCD-Broadcast: definition}
SCD-broadcast consists of 
two operations, denoted $\scdbroadcast()$ and  $\scddeliver()$.
The first operation takes a message to broadcast as input  parameter. 
The second one returns a non-empty set of messages to the process that
invoked it.  By a slight abuse of language, we  say that a process
``scd-delivers a  message $m$'' when it delivers a message set $ms$
containing $m$.

SCD-broadcast is defined by the following
set of properties, where we assume --without loss of generality--
that all the messages that are scd-broadcast are different. 
\begin{itemize}
\vspace{-0.1cm}
\item SCD-Validity.
  If a process scd-delivers a set containing a message $m$,
  then $m$ was scd-broadcast by some process.
\vspace{-0.2cm}
\item SCD-Integrity.
  A message is scd-delivered at most once by each process.
\vspace{-0.2cm}
\item SCD-Ordering.
  If a process $p_i$ scd-delivers first a message $m$ belonging to a set
  $ms_i$ and later a message $m'$ belonging to a set $ms_i'\neq ms_i$, then 
  no process  scd-delivers first the message $m'$ in some
  scd-delivered set $ms'_j$ and later the message $m$ in some scd-delivered
  set $ms_j\neq ms'_j$. 
\vspace{-0.2cm}
\item SCD-Termination-1.
  If a non-faulty process scd-broadcasts a message $m$,
   it terminates its scd-broadcast invocation and
  scd-delivers a message set containing $m$.
  \vspace{-0.2cm}
\item SCD-Termination-2.
  If a process scd-delivers a message set containing $m$,
  every  non-faulty process  scd-delivers a message set containing $m$.
\end{itemize}

\Xomit{\paragraph{A containment property}
Let $ms_i^\ell$ be the  $\ell$-th message set  scd-delivered by $p_i$.
Hence, at some time, $p_i$ scd-delivered the sequence
of message sets $ms_i^1,~\cdots, ms_i^x$.
Let $\MS_i^x= ms_i^1\cup \cdots \cup ms_i^x$. 
The following  property follows directly from the 
SCD-Ordering and Termination-2 properties: 
\begin{itemize}
\vspace{-0.2cm} 
\item SCD-Containment.
$\forall~i,j,x,y$:
$(\MS_i^x \subseteq \MS_j^y) \vee (\MS_j^y\subseteq  \MS_i^x)$.
\end{itemize}
}
\paragraph{$k$-SCD-Broadcast: definition}
This communication abstraction is SCD-Broadcast
strengthened with the following additional property: 
\begin{itemize}
\vspace{-0.1cm}
\item KSCD-Bounded.
  No message set $ms$   kscd-delivered to a process contains more
  than $k$ messages. 
\end{itemize}
In the following, all properties of $k$-SCD-broadcast are prefixed by ``KSCD''. 

\subsection{From $k$-SCD-Broadcast to $k$-BO-Broadcast}

\paragraph{Description of the algorithm}
Algorithm~\ref{algo:kscd-to-kbo} implements  $k$-BO-Broadcast
on top of any system model providing $k$-SCD-Broadcast.
It is an  extremely simple self-explanatory algorithm. 

\begin{algorithm}[h!]
\centering{\fbox{
\begin{minipage}[t]{150mm}
\footnotesize 
\renewcommand{\baselinestretch}{2.5}
\resetline
\begin{tabbing}
aaaaa\=aa\=aaa\=aaaaaa\=\kill

{\bf operation} $\kbobroadcast(v)$ {\bf is} 
$\kscdbroadcast(m)$.~\\~\\

{\bf when a message set} $ms$ {\bf is kscd-delivered do} 
{\bf for each}  $m \in ms$ {\bf do} 
                           $\kbodeliver(m)$   {\bf end for}. 
\end{tabbing}
\end{minipage}
}
\caption{From $k$-SCD-broadcast to  $k$-BO-broadcast}
\label{algo:kscd-to-kbo}
}
\end{algorithm}

\begin{theorem} 
\label{theorem:proof-kscd-to-kbo}
Algorithm~{\em{\ref{algo:kscd-to-kbo}}} implements
 $\mbox{\em \KBO-broadcast}$ in any system model enriched with 
the communication abstraction $\mbox{\em \KSCD-broadcast}$.
\end{theorem}

\begin{proofT}
Properties  
\KBO-Validity,  \KBO-Integrity,  \KBO-Termination-1 and  \KBO-Termination-2
are direct consequences of  their homonym SCD-broadcast properties.

 To prove the additional \KBO-Bounded  property,
 let us consider a message set $\ms$ containing at least $(k+1)$ messages.
  For  each process $p_i$, let  $\fms_i$ (resp. $\lms_i$) denote
  the first (resp. last) message set containing a message in
  $\ms$ received by $p_i$.  Thanks to the KSCD-Ordering property, there
  exist a message $\fm\in \cap_i~ \fms_i$ and a message $\lm\in \cap_i~\lms_i$.
  (Otherwise, we will have messages $m$ and $m'$ such that
  $m \in \fms_i$ $\wedge$ $m \notin \fms_j$ and
  $m' \notin \fms_i$ $\wedge$ $m' \in \fms_j$.)
  
  Let $\ums_i$ denote the union of all the message sets kscd-delivered
  by $p_i$ starting with the set including $\fms_i$ and finishing with
  the set including $\lms_i$.  As, for each process $p_i$, $\ums_i$
  contains at least the $(k+1)$ messages of $\ms$, we have $\fms_i
  \neq \lms_i$. Therefore, we have $\fm \neq \lm$ and $\fm \mapsto
  \lm$.  It follows that $ms$ cannot be an antichain of
  $\mapsto$. Consequently, the antichains of $\mapsto$ cannot contain 
  more than $k$ messages, hence $\width(\mapsto) \leq k$.
  \renewcommand{\toto}{theorem:proof-kscd-to-kbo}
\end{proofT}

\section{From  Repeated $k$-Set Agreement and Snapshot to $k$-SCD-Broadcast}
\label{sec:kSA-snapshot-to-kSCD}

\subsection{The K2S abstraction}

\paragraph{Definition}
The following object, denoted K2S is used by
Algorithm~\ref{algo:rksa-snap-to-kscd} to implement $k$-SCD-broadcast.
``K2S'' stands for $k$-set agreement plus two snapshots.  A K2S object
provides a single operation, denoted ${\ktwospropose}(v)$ that can be invoked
once by each process.  Its output is a set of sets whose size and
elements are constrained by both $k$-set agreement and  the input
size (number of different values proposed by processes).  The output
$sets_i$ of each process $p_i$ is a non-empty set of non-empty sets,
called views and denoted $view$,  satisfying the following properties. Let
$inputs$ denote the set of different input values proposed by the processes.

\begin{itemize}
\vspace{-0.2cm}
\item K2S-Validity.   $\forall~i$: $\forall~ view \in sets_i$:
  $(m\in view)\Rightarrow$  ($m$ was k2s-proposed by a process).
\vspace{-0.2cm}
\item
  Set Size.  $\forall~i$: $1 \leq |sets_i| \leq \mmin(k,|inputs|)$.
  \vspace{-0.2cm}
\item View Size.   $\forall~i:~ \forall~ view \in sets_i$:
  $(1 \leq |view| \leq \mmin(k,|inputs|))$.
\vspace{-0.2cm}
\item Intra-process Inclusion.$\forall~i:~  \forall~
  view1, view2 \in sets_i$:  $view1 \subseteq
  view2 \vee view2 \subseteq view1$.
\vspace{-0.2cm}
\item Inter-process Inclusion. $\forall~i, j$:
 $sets_i \subseteq sets_j \vee sets_j \subseteq sets_i$.
\vspace{-0.2cm}
\item K2S-Termination.
  If a non-faulty process $p_i$ invokes ${\ktwospropose}()$, it returns
  a set $sets_i$. 
\end{itemize}

\paragraph{Algorithm}
Algorithm~\ref{algo:K2S} implements a K2S object.
It uses an underlying $k$-set agreement object $\KSET$,
and two one-shot snapshot objects denoted $\SNAP1$ and $\SNAP2$. 
The algorithm is a three-phase algorithm. 
\begin{itemize}
\vspace{-0.2cm}
\item
Phase 1 (line~\ref{ksss-01}).
When a process $p_i$ invokes ${\ktwospropose}(v)$, it first
proposes $v$ to the $k$-set agreement object, from which it obtains
a value $val_i$ (line~\ref{ksss-01}).
\vspace{-0.2cm}
\item
Phase 2 (lines~\ref{ksss-02}-\ref{ksss-03}).
Then $p_i$ writes $val_i$ in the
first snapshot object $\SNAP1$,   reads its content, saves it in $snap1_i$,
and computes the set of values ($view_i$) that, from its point of view,
have been proposed to the  $k$-set agreement object.
\vspace{-0.2cm}
\item
Phase 3 (lines~\ref{ksss-04}-\ref{ksss-06}).
Process $p_i$ then writes its view $view_i$ in the second snapshot object
$\SNAP2$, reads its value, and computes the set of views ($sets_i$) 
obtained --as far as it knows-- by the other processes. 
Process $p_i$ finally returns this set of views $sets_i$. 
\end{itemize}

\begin{algorithm}[h!]
\centering{\fbox{
\begin{minipage}[t]{150mm}
\footnotesize 
\renewcommand{\baselinestretch}{2.5}
\resetline
\begin{tabbing}
aaaaa\=aa\=aaa\=aaaaaa\=\kill

{\bf operation} ${\sf \ktwospropose}(v)$ {\bf is}\\

\line{ksss-01} \> $val_i \leftarrow \KSET.{\sf propose}(v)$;\\

\line{ksss-02} \> $\SNAP1.\wwrite(val_i)$;
                  $snap1_i  \leftarrow \SNAP1.{\sf snapshot}()$;\\

\line{ksss-03} \> $view_i  \leftarrow \{snap1_i[j]~|~snap1_i[j] \neq \bot\}$;\\

\line{ksss-04} \> $\SNAP2.\wwrite(view_i)$;
                  $snap2_i  \leftarrow \SNAP2.{\sf snapshot}()$;\\

\line{ksss-05} \>  $sets_i\leftarrow \{snap2_i[j]~|~snap2_i[j] \neq\bot\}$;\\

\line{ksss-06} \> $\return(sets_i)$.

\end{tabbing}
\end{minipage}
}
\caption{An implementation of a K2S object}
\label{algo:K2S}
}
\end{algorithm}

\begin{theorem}
\label{theorem:K2S}
Algorithm~{\em \ref{algo:K2S}} satisfied the properties defining a
{\em K2S} object. 
\end{theorem}  

\begin{proofT}
The K2S-Validity property follows from the $k$-set Validity property, and
the fact that a snapshot object does not modify the values that are written. 
Similarly, the K2S-Termination property follows the Termination properties of
the $k$-set agreement and snapshot objects.

The Intra-process Inclusion property follows from the Containment property
of the views returned from the snapshot object $\SNAP1$ (line~\ref{ksss-02}). 
Similarly, the Inter-process Inclusion property follows from the Containment
property of the sets returned from the snapshot object $\SNAP2$
(line~\ref{ksss-04}).

The fact that no view contains more than  $\mmin(k,|inputs|)$
elements follows from the $k$-set agreement object $\KSET$
which returns at most $k$ different values. The View Size property is
an immediate consequence of this observation.
Finally, the Set Size property follows from the fact that there are at
most  $\mmin(k,|inputs|)$ different views obtained by the processes, and
these views satisfy the Containment property. 
\renewcommand{\toto}{theorem:K2S} 
\end{proofT}
  
\paragraph{Repeated K2S}
In the following we consider a repeated K2S object, denoted $\KSS$.
A process $p_i$  invokes 
$\KSS.{\ktwospropose}(r,v)$ where $v$ is the value it proposes to
the instance number $r$. The instance numbers used by each process are
increasing (but not necessarily consecutive).  Hence, two snapshot objects
are associated with every K2S instance, and  line~\ref{ksss-01} of
Algorithm~\ref{algo:K2S} becomes  $\KSET.{\propose}(r,v)$.

\subsection{From $k$-Set Agreement and Snapshot to $k$-SCD-Broadcast}

Algorithm~\ref{algo:rksa-snap-to-kscd} builds the $k$-SCD-Broadcast
abstraction on top  $k$-set agreement and snapshot objects.

\paragraph{Shared objects and local objects}
\begin{itemize}
\vspace{-0.2cm}
\item The processes cooperate through two concurrent objects:
$\MEM[1..n]$, a multishot snapshot object, such that 
  $\MEM[i]$ contains the set of messages kscd-broadcast by $p_i$,
  and a repeated K2S object denoted  $\KSS$. 
\vspace{-0.2cm}
\item
A process $p_i$ manages two local copies of $\MEM$ denoted
$mem1_i$ and $mem2_i$, two auxiliary sets 
$\todeliver1_i$ and $\todeliver2_i$,  and a set $delivered_i$, which
contains all the messages it has locally kscd-delivered; 
$\mem1_i[i]$ is initialized to an empty set. 
\vspace{-0.2cm}
\item $r_i$  denotes the next round number
  that $p_i$ will execute;  $sets_i$ is a local set whose aim is to
  contain the set of message sets returned by the last invocation of a
  K2S object.
\vspace{-0.2cm}
\item  
Each process $p_i$ manages two sequences of messages sets, both initialized
to  $\epsilon$ (empty sequence), denoted  $seq_i$ and $new\_seq_i$;
$\head(sq)$ returns the first element of the sequence $sq$,
and $\tail(sq)$ returns the sequence $sq$ without its first element;
$\oplus$ denotes sequence concatenation.

The aim of the local sequence $new\_seq_i$ is to contain a sequence
of message sets obtained from $sets_i$ (last invocation of a K2S object)
such that no message belongs to several message sets. 
 
As far as $seq_i$ is concerned, we have the following
(at line~\ref{nRW-19} of Algorithm~\ref{algo:rksa-snap-to-kscd}).
Let $seq_i= ms_1,~ms_2,~\cdots, ms_\ell$, where $1\leq \ell\leq k$ and 
each $ms_x$ is a message set. 
This sequence can be decomposed into two (possibly empty) sub-sequences
$ms_1,~ms_2,~\cdots, ms_y$ and $ms_{y+1} \cdots,ms_\ell$ such that:
\begin{itemize}
\vspace{-0.2cm}
\item
$ms_1,~ms_2,~\cdots, ms_y$ can be in turn decomposed as follows:\\
$(ms_1\cup ms_2\cup \cdots\cup ms_a), (ms_{a+1}\cup ms_{a+2}\cup \cdots\cup ms_b),
~\cdots, (ms_c\cup \cdots \cup ms_y)$ \\
where each union set (e.g., $ms_{a+1}\cup ms_{a+2}\cup \cdots\cup ms_b$) 
is a message set that has been kscd-delivered by some process 
(some  union sets can contain a single message set)\footnote{Let us remark that
  it is possible that, while a process kscd-delivered the message set
  $ms= ms_1\cup ms_2\cup \cdots\cup ms_a$, another process
  kscd-delivered the messages in  $ms$ in several messages sets, e.g., 
  first the message set $ms_1\cup ms_2\cup ms_3$ and
  then the message set $ms_4\cup \cdots\cup ms_a$.}.
\vspace{-0.1cm}
\item For each $x:~ y+1\leq x\leq \ell:~ m_x$ is a message set whose  
 messages have not yet been  kscd-delivered by a process.
\end{itemize}
\end{itemize}

\paragraph{Operation $\kscdbroadcast()$}
When it invokes  $\kscdbroadcast()$, a process $p_i$  first
adds $m$ to the shared memory $\MEM$, which contains all the
messages it has already kscd-broadcast (line~\ref{nRW-01}). 
Then $p_i$ reads atomically the whole content of $\MEM$, which is saved
in $mem1_i$ (line~\ref{nRW-01}). Then, $p_i$  computes the set of messages
not yet locally kscd-delivered  and waits until
all these messages appear in kscd-delivered message sets (line~\ref{nRW-02}). 
Let us notice that, it follows from these statements, that a process has
kscd-delivered its previous message when it issues its next $\kscdbroadcast()$.

\begin{algorithm}[h!]
\centering{\fbox{
\begin{minipage}[t]{150mm}
\footnotesize 
\renewcommand{\baselinestretch}{2.5}
\resetline
\begin{tabbing}
aaaaa\=aa\=aaa\=aaaaaa\=\kill

{\bf operation} $\kscdbroadcast(m)$ {\bf is}\\

\line{nRW-01} \> $\MEM.\wwrite(\mem1_i[i] \cup \{m\})$;
                 $\mem1_i \leftarrow \MEM.\snapshot()$;\\

\line{nRW-02} \> $\todeliver1_i \leftarrow
  (\cup_{1\leq j \leq n}~\mem1_i[j]) \setminus \mathit{delivered}_i$;

${\wait}(\todeliver1_i \subseteq \mathit{delivered}_i)$.~\\~\\

{\bf background task} $T$ {\bf is}\\

\line{nRW-03} \> {\bf repeat forever}\\

\line{nRW-04} \>\> $\mathit{prop}_i \leftarrow \bot$;\\

\line{nRW-05} \>\>  {\bf if} \=$(\mathit{seq}_i = \epsilon)$ \= 

      {\bf then}  \= $\mem2_i \leftarrow \MEM.\snapshot()$;\\

\line{nRW-06} \>\> \> \>  \> $\todeliver2_i \leftarrow
  (\cup_{1\leq j \leq n}~\mem2_i[j]) \setminus \mathit{delivered}_i$;\\

\line{nRW-07} \>\> \>  \> \> {\bf if} \=$(\todeliver2_i \neq \emptyset)$
     {\bf then}
       $\mathit{prop}_i \leftarrow \textrm{ a message }\in \todeliver2_i$
           {\bf end if}\\ 

\line{nRW-08} \>\> \> \>  {\bf else}
\> $\mathit{prop}_i \leftarrow
           \textrm{a message of the first message set of }\mathit{seq}_i$\\

\line{nRW-09} \>\>  {\bf end if};\\

\line{nRW-10} \>\> {\bf if}  $(\mathit{prop}_i \neq\bot)$ \\

\line{nRW-11} \>\>\>
     {\bf then} \= $r_i \leftarrow |\mathit{delivered}_i|$;

 $\mathit{sets}_i \leftarrow \KSS.{\ktwospropose}(r_i,\mathit{prop}_i)$;
     $new\_seq_i \leftarrow \epsilon$;\\

\line{nRW-12} \>\>\>\> {\bf while} \=  $(sets_i \neq \{\emptyset\})$ {\bf do}\\

\line{nRW-13} \>\> \> \> \> 
$\mathit{min\_set}_i\leftarrow$ 
              non-empty set of minimal size in $\mathit{sets}_i$;\\

\line{nRW-14} \>\> \> \> \> $\mathit{new\_seq}_i
               \leftarrow \mathit{new\_seq}_i   \oplus \mathit{min\_set}_i$;\\

\line{nRW-15} \>\> \> \> \> {\bf for each set} $s\in sets_i$ {\bf do}
            $sets_i \leftarrow (sets_i \setminus \{s\})
                \cup \{\mathit{s \setminus min\_set}_i\}$ {\bf end for}\\

\line{nRW-16} \>\> \> \> {\bf end while};\\

\line{nRW-17} \>\>\>\> 

{\bf let} $aux_i$ = all the messages in the sets of  $\mathit{new\_seq}_i$;\\

\line{nRW-18} \>\>\>\>  {\bf for each set } $s\in seq_i$ {\bf do}
                          $s \leftarrow s \setminus aux_i$ {\bf end for};\\

\line{nRW-19} \>\>\>\>
      $\mathit{seq}_i \leftarrow \mathit{new\_seq}_i   \oplus  \mathit{seq}_i$;
     {\bf let} $\first_i=\head(seq_i)$;  {\bf let} $\rest_i=\tail(seq_i)$;  \\

\line{nRW-20} \>\>\>\>
       $\kscddeliver(\first_i)$; $\mathit{delivered}_i \leftarrow
                                    \mathit{delivered}_i  \cup \first_i$;
       $\mathit{seq}_i \leftarrow \rest_i$\\

\line{nRW-21} \>\>   {\bf end if}\\ 

\line{nRW-22} \> {\bf end repeat}.

\end{tabbing}
\end{minipage}
}
  \caption{From $k$-set agreement  and snapshot
    objects to  $k$-SCD-broadcast (code for $p_i$)}
\label{algo:rksa-snap-to-kscd}
}
\end{algorithm}

\paragraph{Underlying task $T$}
This task  is the core of the algorithm. It consists of an infinite
loop, which implements a sequence of asynchronous rounds
(lines~\ref{nRW-11}-\ref{nRW-20}).  Each process $p_i$ executes a
sub-sequence of non-necessarily consecutive rounds. Moreover, any two
processes do not necessarily execute the same sub-sequence of
rounds. The current round of a process $p_i$ is defined by the value of
$|\delivered_i|$ (number of messages already locally kscd-delivered).

The progress of a process from a round $r$ to its next round
$r'>r$ depends on the size of the message set (denoted $\first_i$
in the algorithm, line~\ref{nRW-20}) it kscd-delivers at the end of
round $r$ ($\delivered_i$ becomes then $\delivered_i \cup \first_i$).
The message set $\first_i$ depends on the values returned by the
K2S object associated with the round $r$, as explained below.

\paragraph{Underlying task $T$: proposal computation}
(Lines~\ref{nRW-04}-\ref{nRW-09}) Two rounds executed by a process
$p_i$ are separated by the local computation of a message value
($prop_i$) that $p_i$ will propose to the next K2S object.
This local computation is as follows (lines~\ref{nRW-05}-\ref{nRW-09}),
where $seq_i$ (computed at lines~\ref{nRW-18}-\ref{nRW-20}) is a sequence
of message sets that, after some ``cleaning'', are candidates to be
locally kscd-delivered.  There are two cases.
\begin{itemize}
\vspace{-0.2cm}
\item
  Case 1: $seq_i=\emptyset$.
In this case (similarly to line~\ref{nRW-02}) $p_i$
computes the set of messages ($to\_deliver2_i$)  it sees as kscd-broadcast
but not yet locally kscd-delivered (lines~\ref{nRW-05}-\ref{nRW-06}).
If  $to\_deliver2_i\neq \emptyset$, a message of this set becomes its
proposal $prop_i$ for the K2S object associated with the next round
(line~\ref{nRW-07}).
Otherwise, we have $prop_i=\bot$,
which, due to the predicate of line~\ref{nRW-10}, entails
a new execution of the loop (skipping lines~\ref{nRW-11}-\ref{nRW-20}).
\vspace{-0.2cm}
\item
Case 2: $seq_i\neq\emptyset$. In this case, $prop_i$ is assigned a
message of the first message set of $seq_i$ (line~\ref{nRW-08}).
\end{itemize}

\paragraph{Underlying task $T$: benefiting from a K2S object to
kscd-deliver a message set} (Lines~\ref{nRW-11}-\ref{nRW-20}) If a
proposal has been previously computed (predicate of
line~\ref{nRW-10}), $p_i$ executes its next round, whose number is
$r_i=|\delivered_i|$. The increase step of $|\delivered_i|$ can vary from
round to round, and can be any value $\ell\in[1..k]$,
lines~\ref{nRW-14} and~\ref{nRW-15}).  As already indicated, while the
round numbers have a global meaning (the same global sequence of
rounds is shared by all processes), each process executes  a subset of
this sequence (as defined by the increasing successive values of
$|delivered_i|$).  Despite the fact processes skip/execute different
round numbers, once combined with the use of K2R objects, round
numbers allow processes to synchronize in a consistent way.
This round synchronization property is captured by
Lemmas~\ref{lemma:same-round-same-msgs}-\ref{lemma:same-msgs}.

From an operational point of view, a round starts with the invocation
$\KSS.{\ktwospropose}(r_i,\mathit{prop}_i)$ where
$r_i=|\delivered_i|$, which returns a set of message sets
$\mathit{sets}_i$ (line~\ref{nRW-11}).  Then (``while'' loop at
lines~\ref{nRW-12}-\ref{nRW-16}), $p_i$ builds from the message sets
belonging to $sets_i$ a sequence of message sets $new\_seq_i$, that
will be used to extract the next message set kscd-delivered by $p_i$
(lines~\ref{nRW-17}-\ref{nRW-20}).
The construction of $new\_seq_i$ is as follows. Iteratively, $p_i$
takes the smallest set of $sets_i$ ($min\_set_i$, line~\ref{nRW-13}),
adds it at the end of $new\_seq_i$ (line~\ref{nRW-14}), and purges all
the sets of $sets_i$ from the messages in $min\_set_i$ (line~\ref{nRW-15}),
so that no message will locally appear in two different messages sets
of $new\_seq_i$.

When $new\_seq_i$ is built, $p_i$ first purges all the sets of the sequence
$seq_i$ from the messages in $new\_seq_i$ (lines~\ref{nRW-17}-\ref{nRW-18}),
and adds then $new\_seq_i$ at the front  of $seq_i$ (line~\ref{nRW-19}).
Finally, $p_i$ kscd-delivers the first message set of $seq_i$, and
updates accordingly $delivered_i$ and $seq_i$  (lines~\ref{nRW-20}).

\subsection{Proof of the algorithm}

\begin{lemma}
\label{lemma-kSCD-bounded}
A message set kscd-delivered (line~{\em{\ref{nRW-20}}})
contains at most $k$ messages. 
\end{lemma}

\begin{proofL}
Let us consider a process $p_i$ that executes $\kscddeliver(\first_i)$
at line~\ref{nRW-20}.  The set $\first_i$ is the first message set of
$new\_seq_i$ computed at lines~\ref{nRW-12}-\ref{nRW-16}. More precisely, 
it is the first set $min\_set_i$ computed by $p_i$ at line~\ref{nRW-13},
which means it is a set belonging to $sets_i$, the set of message sets 
locally returned by the invocation of
$\KSS.{\ktwospropose}(r_i,\mathit{prop}_i)$ at line~\ref{nRW-11}.
It then follows from the View Size property of
the message sets returned by $\KSS.{\ktwospropose}(r_i,-)$
that $\first_i$ contains at most $k$ messages.   
\renewcommand{\toto}{lemma-kSCD-bounded} 
\end{proofL}

\begin{lemma}
 \label{lemma-kSCD-Validity}
 If a process kscd-delivers a message set containing
 a message $m$,  $m$ was kscd-broadcast by a process.
\end{lemma}

\begin{proofL}
Let a message $m\in \first_i$, which is kscd-delivered by a process
$p_i$ during a round $r$. It follows (as seen in the proof of the
previous lemma) that $m$ was proposed by a process $p_j$ that invoked 
$\KSS.{\ktwospropose}(r,\mathit{prop}_j)$ at line~\ref{nRW-11}.
As $prop_j\neq\bot$ (line~\ref{nRW-10}), $prop_j$ was assigned $m$
at line~\ref{nRW-07} or at line~\ref{nRW-08}. There are two cases.
\begin{itemize}
\vspace{-0.2cm}
\item 
$prop_j$ was assigned at line~\ref{nRW-07}. It then follows from 
lines~\ref{nRW-05}-\ref{nRW-06} that $m$ was written in the snapshot object
$\MEM$. As this can occurs only at line~\ref{nRW-01}, it follows that
$m$ was written into $\MEM$ by a process that invoked  $\kscdbroadcast(m)$,
which proves the lemma. 
\vspace{-0.2cm}
\item $prop_j$ was assigned at line~\ref{nRW-08}. In this case,
$seq_j\neq\epsilon$ and $m$ is a message in the first message set of
$seq_j$. We claim that $seq_i$ contains message sets that contain only
messages that have been kscd-broadcast. The proof of the lemma then follows. 

Proof of the claim.
New message sets ($new\_seq_j$) are added to $seq_j$ only at
line~\ref{nRW-19}, and sequences $new\_seq_j$ are built only
at line~\ref{nRW-14}, with  messages sets (purged not to have
two message sets including a same message) obtained from invocations
by $p_j$ of  ${\ktwospropose}()$ on K2S objects.
These messages sets include only messages proposed to these objects
(K2S-Validity property). These messages  (values $prop_x$ proposed by
processes $p_x$) come from line~\ref{nRW-07} or~\ref{nRW-08}. 
If $m=prop_x$ was computed at line~\ref{nRW-07}, it follows from the
previous item that it was kscd-broadcast by some process. 
If $m=prop_x$ was computed by $p_x$ at line~\ref{nRW-08}, it comes from
$seq_x$. In this case, the proof follows from a simple induction argument,
starting from  round $0$, which concludes the proof of the claim. 
\end{itemize}
\vspace{-0.7cm}  
\renewcommand{\toto}{lemma-kSCD-Validity} 
\end{proofL}

\vspace{-0.5cm}
\paragraph{Notations}
\begin{itemize}
\vspace{-0.2cm}
\item 
 $msg\_set_i(r)$ =  message set kscd-delivered by process $p_i$ at
round $r$ if $p_i$ participated in it, and $\emptyset$ otherwise.
\vspace{-0.2cm}
\item 
$seq_i(r)$ =  value of $seq_i$ at the end of the last round $r'\leq r$
in which $p_i$ participated.
\vspace{-0.2cm}
\item 
$msgs_i(r,r')$ = set of
messages contained in message sets kscd-delivered  by $p_i$ between round $r$
(included) and round $r'>r$ (not included), i.e. $msgs_i(r,r')
= \bigcup_{r \leq r'' < r'} msg\_set_i(r'')$.
\vspace{-0.2cm}
\item 
$\KSS(r)$ = K2S instance accessed by $\KSS.{\ktwospropose}(r,-)$
(line~\ref{nRW-11}).
\vspace{-0.2cm}
\item 
 $sets_i(r)$ = set of message sets  obtained by $p_i$ from $\KSS[r]$.
\end{itemize}

\begin{lemma} 
\label{lemma:same-round-pref-included}
Let $p_i$ and $p_j$ be two  processes that terminate  round 
$r$, with $|msg\_set_i(r)| \leq |msg\_set_j(r)|$. Then {\em (i)} 
$msg\_set_i(r) \subseteq msg\_set_j(r)$, and  {\em (ii)} there is a prefix
$\prefi$ of $seq_i(r)$ such that $msg\_set_j(r)= msg\_set_i(r)
\cup (\bigcup_{msg\_set ~\in~ \prefi} msg\_set)$.
\end{lemma}

\begin{proofL}
Let $p_i$ and $p_j$ be two  processes that kscd-deliver  the 
message sets $msg\_set_i(r)$ and $msg\_set_j(r)$,  respectively, 
these sets being  such that  $|msg\_set_i(r)| \leq |msg\_set_j(r)|$.
Let us observe that,  as both $p_i$ and $p_j$ invoked
$\KSS.{\ktwospropose}(r,-)$ (lines~\ref{nRW-11} and~\ref{nRW-20}),
we have  $sets_i(r) \subseteq sets_j(r)$ or
$sets_j(r) \subseteq sets_i(r)$ (Inter-process Inclusion). 

As  $|msg\_set_i(r)| \leq |msg\_set_j(r)|$, it follows from the
Inter-process and Intra-process inclusion properties of $\KSS(r)$,
and the definition of $msg\_set_i(r)=\first_i=min\_set_i\in sets_i(r)$,
and $msg\_set_j(r)=\first_j=min\_set_j\in sets_j(r)\subseteq sets_i(r)$, that
$msg\_set_i(r) \subseteq msg\_set_j(r)$, which completes the proof of (i).

As far as (ii) is concerned, we have the following.
If $msg\_set_i(r) = msg\_set_j(r)$, we have $\prefi=\epsilon$ and the
lemma follows. So, let us assume $msg\_set_i(r) \subsetneq msg\_set_j(r)$.
As $msg\_set_i(r)$ is the smallest message set of $sets_i(r)$
(lines~\ref{nRW-13}-\ref{nRW-14}  and~\ref{nRW-19}-\ref{nRW-20}), 
and $msg\_set_j(r)$ is the smallest message set of $sets_j(r)$, 
it follows that $sets_j(r) \subset sets_i(r)$.  The property
$msg\_set_j(r) = msg\_set_i(r)  \cup (\bigcup_{msg\_set ~\in~ \prefi} msg\_set)$
follows then from the following observation.
Let  $sets_i(r)=\{s_1,s_2,...,s_\ell\}$, where $\ell \leq  k$ and 
$s_1\subsetneq s_2 \subsetneq \cdots \subsetneq s_\ell$.
As $sets_j(r) \subset sets_i(r)$, one $s_x$ is $msg\_set_j(r)$. 
It follows that the union of the  sets $min\_set_i$ computed by $p_i$
in the while loop of round $r$ (lines~\ref{nRW-13}-\ref{nRW-15})
eventually includes all the  messages of $msg\_set_j(r)$,
from which we conclude that there is a prefix $\prefi$ of $seq_i(r)$
(lines~\ref{nRW-12}-\ref{nRW-19}, namely a prefix of the sequence
$new\_seq_i$, which is defined from the sequence
of the sets $min\_set_i$), such that
$msg\_set_j(r)= msg\_set_i(r) \cup (\bigcup_{msg\_set ~\in~ \prefi} msg\_set)$,
which completes the proof of the lemma. 
\renewcommand{\toto}{lemma:same-round-pref-included}
\end{proofL}

\noindent
The next two lemmas capture the global message set delivery synchronization
among the processes.
\begin{lemma}
\label{lemma:same-round-same-msgs}
Let $p_i$ and $p_j$ be two  processes that terminate round
$r' \geq r+|msg\_set_j(r)|$, and are such that
$|msg\_set_i(r)| \leq |msg\_set_j(r)|$. Then
 {\em(i)} $msgs_i(r,r+|msg\_set_j(r)|) = msgs_j(r,r+|msg\_set_j(r)|)$, and
 {\em(ii)} $p_i$ and $p_j$ will both participate in round $r+|msg\_set_j(r)|$.
\end{lemma}

\begin{proofL}
If $|msg\_set_i(r)| = |msg\_set_j(r)| =\alpha$, both $p_i$ and $p_j$
are such that $|\delivered_i| = |\delivered_j|= r + \alpha$
when they terminate round $r$. Consequently,  they both 
proceed from round $r$ to round $r+\alpha$, thereby skipping the rounds
from $r+1$ until $r+\alpha-1$.  We  then have (i)
$msgs_i(r,r+|msg\_set_j(r)|) =msg\_set_i(r)=
msg\_set_j(r)= msgs_j(r,r+|msg\_set_j(r)|)$, 
(ii) both $p_i$ and $p_j$ will participate in round $r+|msg\_set_j(r)|$,
and the lemma follows.

Hence, let us consider that $|msg\_set_i(r)|=\alpha <
|msg\_set_j(r)|=\alpha +\beta$.  The next round executed by $p_i$ will
be the round $r+\alpha$, while the next round executed by $p_j$ will
be the round $r+\alpha+\beta$.  Moreover, to simplify and without loss
of generality, let us assume that $msg\_set_i(r)$
(resp. $msg\_set_j(r)$) is the smallest 
(resp. second smallest) message set
in the sets of message sets $sets$ output by $\KSS(r)$.

According to Lemma \ref{lemma:same-round-pref-included}, after round
$r$, the first element of $seq_i$ is $msg\_set_j(r) \setminus
msg\_set_i(r)$.  This also applies to any other process that delivered
$msg\_set_i(r)$ at round $r$. At round $r+\alpha$, all these processes
will then propose a message in $msg\_set_j(r) \setminus
msg\_set_i(r)$.  Because of the K2S-Validity property of
$\KSS(r+\alpha)$, all these processes will then deliver a subset of
$msg\_set_j(r) \setminus msg\_set_i(r)$.  For the same reason, until
round $r+\alpha+\beta$, no process will propose a message not in
$msg\_set_j(r) \setminus msg\_set_i(r)$. At round $r+\alpha+\beta$,
they will then have delivered all the messages in $msg\_set_j(r)
\setminus msg\_set_i(r)$, and they will participate in round
$r+\alpha+\beta$, from which the lemma follows.
\renewcommand{\toto}{lemma:same-round-same-msgs}
\end{proofL}

\begin{lemma}
\label{lemma:same-msgs}
Let $r$ be a round in which all the non-faulty processes participate.
There is a round $r'$ with $r < r' \leq r+k$ in which all non-faulty
processes participate and such that, for any pair of non-faulty processes
$p_i$ and $p_j$, we have  $msgs_i(r,r') = msgs_j(r,r')$.
\end{lemma}

\begin{proofL}
  As initially $\forall~i:~|\delivered_i|=0$, all the non-crashed
processes invoke $\KSS.{\ktwospropose}(0,-)$.  We prove that there is
a round $r\in[1..k]$ in which all the non-crashed processes
participate, and for any pair of them $p_i$ and $p_j$, we have
$msgs_i(0,r) =msgs_j(0,r)$. This constitute the base case of an
induction.  Then, the same reasoning can be used to show that if the
non-faulty processes participate in a round $r$, there is a round $r'$
with $r < r' \leq r+k$  and such that, for any pair of non-faulty
processes $p_i$ and $p_j$, we have $msgs_i(r,r') = msgs_j(r,r')$.

Let us consider any two processes $p_i$ and $p_j$ that terminate round $0$.
Moreover, without loss of generality, let us assume that, among
the sets of message sets output by  $\KSS(0)$, 
$sets_i(0)$ is the greatest and $sets_j(0)$ is the smallest. 
It follows from the Inter-process inclusion property that
$sets_j(0) \subseteq sets_i(0)$, and  from line~\ref{nRW-13} plus the
Intra-process inclusion property that $msg\_set_i(0) \subseteq msg\_set_j(0)$.
Hence, $|msg\_set_i(0)| \leq | msg\_set_j(0)|$. 
Moreover, due to the View size property of  $\KSS(0)$ we have
$|msg\_set_i(0)| \leq | msg\_set_j(0)|=r \leq k$.
Applying Lemma~\ref{lemma:same-round-same-msgs}, we have
$msg_i(0, 0+r)=msg_j(0, 0+r)$, which concludes the proof of the lemma.   
\renewcommand{\toto}{lemma:same-msgs}
\end{proofL}

\begin{lemma}
\label{lemma-kSCD-msg-ordering} 
If a process $p_i$ kscd-delivers first a message $m$ belonging to a set
  $ms_i$ and later a message $m'$ belonging to a set $ms_i'\neq ms_i$, then 
  no process  kscd-delivers first the message $m'$ in some
  kscd-delivered set $ms'_j$ and later the message $m$ in some kscd-delivered
  set $ms_j\neq ms'_j$. 
\end{lemma}

\begin{proofL}
Let us first note that, at each process, the kscd-delivery of message
sets establishes a partial order on messages. Given a process $p_i$,
let $\rightarrow_i$ be the partial order defined as
follows\footnote{This definition is similar to the definition of
  $\mapsto_i$ given in Section~\ref{sec:kBO} devoted to
  $k$BO-broadcast.}: $m \rightarrow_i m'$ if $p_i$ kscd-delivered
first a message set $ms_i$ including $m$, and later kscd-delivered a
message set $ms'_i$ including $m'$.  Hence, if $m$ and $m'$ were
kscd-delivered in the same message set by $p_i$, we have $m
\not\rightarrow_i m'$ and $m' \not\rightarrow_i m$.  

Let us also note that, along the execution of a process $p_i$, the partial 
order $\rightarrow_i$ can only be extended, i.e. if $m \rightarrow_i m'$ 
at time $t$, we cannot have $m \not\rightarrow_i m'$ at time $t'>t$. 
This, along with the fact that a faulty process executes its algorithm 
correctly until it crashes, allows us to consider, in the context of 
this proof, that $p_i$ and $p_j$ are non-faulty.

In order to prove the lemma, we then have to show that the partial
orders $\rightarrow_i$ and $\rightarrow_j$ are compatible, i.e. for
any two messages $m$ and $m'$, $(m\rightarrow_i m') \Rightarrow
(m'\not\rightarrow_j m)$ and $(m\rightarrow_j m') \Rightarrow
(m'\not\rightarrow_i m)$.

According to Lemma \ref{lemma:same-msgs}, for each round $r$ in which
all processes participate, there is a round $r' > r$ in which all
processes participate. Moreover, for any two non-faulty process $p_i$
and $p_j$, we have $msgs_i(r,r')=msgs_j(r,r')$. For any such round
$r$, we then have that if $p_i$ delivered message $m$ strictly before
round $r$ and delivered $m'$ at round $r$ or afterwards, we have both
$(m\rightarrow_i m')$ and $(m'\rightarrow_j m)$.  We will then
consider the messages delivered between two such rounds $r$ and $r'$.

Without loss of generality, suppose that the message set
kscd-delivered by $p_i$ at round $r$ is smaller than, or equal to, the
message set kscd-delivered by $p_j$ at the same round, i.e.
$|msg\_set_i(r)| \leq |msg\_set_j(r)|$.  It follows from
Lemma~\ref{lemma:same-round-same-msgs} that $msgs_i(r,|msg\_set_j(r)|)
= msgs_j(r,|msg\_set_j(r)|)$. Moreover, as all the messages in
$msg\_set_j(r)$ were kscd-delivered by $p_j$ in a single set, they are
all incomparable when considering $\rightarrow_j$.  The partial orders
$\rightarrow_i$ and $\rightarrow_j$, when restricted to the messages
in $msg\_set_j(r)$, are thus compatible.

According to  Lemma \ref{lemma:same-round-same-msgs},
$p_i$ and $p_j$ will both participate in round $r+\alpha = r+|msg\_set_j(r)|$. 
If $r+\alpha = r'$, the lemma follows. 
Otherwise, let $\beta =
\max(|msg\_set_i(r+\alpha)|,|msg\_set_j(r+\alpha)|)$.  The previous
reasoning, again due to Lemma \ref{lemma:same-round-same-msgs}, can
then be applied again to the messages in
$msgs_i(r+\alpha,r+\alpha+\beta) = msgs_j(r+\alpha,r+\alpha+\beta)$,
and $p_i$ and $p_j$ will both participate in round $r+\alpha + \beta$.
This can be repeated until round $r'$, showing that the partial orders
$\rightarrow_i$ and $\rightarrow_j$ are compatible, which concludes
the proof of the lemma.  \renewcommand{\toto}{lemma-kSCD-msg-ordering}
\end{proofL}

\begin{lemma}
\label{lemma:kSCD-no-duplication}
No message $m$ is kscd-delivered twice by a process $p_i$.
\end{lemma}

\begin{proofL}
Let us consider a sequence of message sets $seq_i$. 
\begin{itemize}
\vspace{-0.2cm}
\item
Due to line~\ref{nRW-15} (update of $sets_i$)
and lines~\ref{nRW-17}-\ref{nRW-18} (update of $seq_i$), 
no message can appear twice in the message sets of $seq_i$.
\vspace{-0.2cm}
\item
Due to the predicate of line~\ref{nRW-05} (when $seq_i\neq\epsilon$),
line~\ref{nRW-08} (assignment of $prop_i$), 
and line~\ref{nRW-20} (updates of $\delivered_i$ and $seq_i$),
all messages of $seq_i$ are kscd-delivered are added to $\delivered_i$
at their kscd-delivery time. 
\vspace{-0.2cm}
\item
When $seq_i$ becomes empty at line~\ref{nRW-20},
due to the previous  update of $\delivered_i$ at the same line,
and the update of  $\todeliver2_i$ at line~\ref{nRW-06}, it follows
that no message already kscd-delivered  can appear in $\todeliver2_i$.   
\end{itemize}
The lemma follows from the previous observations. 
\renewcommand{\toto}{lemma:kSCD-no-duplication}
\end{proofL}

\begin{lemma}
\label{lemma-kSCD-no-blocking}
Let $m$ be a message that has been deposited into $\MEM$.
Eventually, $m$ is kscd-delivered (at least) by the  non-faulty processes.
\end{lemma}

\begin{proofL}
Let  $m$ be a message that has been deposited in the snapshot
object $\MEM$ by some  process  $p_i$.
Then,  due to the definition of  $\todeliver1_i$ (line~\ref{nRW-02})
and the predicate at the same line,   $p_i$ cannot
kscd-broadcast another message before it has  kscd-delivered $m$,
and any $p_j\neq p_i$ can kscd-broadcast at most one message before
it kscd-delivers $m$.
Hence, considering any process $p_x$, it follows that its set
$\todeliver1_x$ can contain at most $n$ messages, and $p_x$ is
prevented from kscd-broadcasting new messages before all the messages
in $\todeliver1_x$ have been kscd-delivered.

Let us assume by contradiction that a non-faulty process $p_x$ 
never kscd-delivers a message set containing $m$.
Either because $seq_x\neq\epsilon$ or $\todeliver2_x\neq \emptyset$, 
$p_x$ computes a value $prop_x$ and invokes
$\KSS.{\ktwospropose}(|\delivered_x|,prop_i)$, from which it obtains a
set of message sets $sets_x$.
Process $p_x$ then kscd-delivers message sets extracted
from $sets_x$ at lines~\ref{nRW-11}-\ref{nRW-20}. 
It follows that the set $\delivered_x$ increases (line~\ref{nRW-20}). 
This can occur a finite number of times, after which the only message
in $\todeliver1_x\setminus \delivered_i$ and
$\todeliver2_x\setminus \delivered_i$ is $m$.

The previous observation is true for all the processes that have not
yet crashed, from which it follows that there is a finite time after
which the only value $prop_x$ that can be proposed by a process to a
K2S object is $m$. It follows that only the set of message sets
$\{\{m\}\}$ can be output by such an object. It  follows that $p_x$
kscd-delivers the message set $\{m\}$, from which we conclude that at least
all the non-faulty processes kscd-deliver a message set containing $m$.
\renewcommand{\toto}{lemma-kSCD-no-blocking}
\end{proofL}

\begin{lemma}
\label{lemma-kSCD-termination-2} 
If a process kscd-delivers a message $m$, every non-faulty process
kscd-delivers a message set  containing $m$.
\end{lemma}

\begin{proofL}
If a process kscd-delivers a message set containing a message $m$, 
this message was added to $\MEM$ by some process at line~\ref{nRW-01}. 
The lemma then follows from  Lemma~\ref{lemma-kSCD-no-blocking}.  
\renewcommand{\toto}{lemma-kSCD-termination-2} 
\end{proofL}

\begin{lemma}
\label{lemma-kSCD-termination-1}
If a non-faulty process $p_i$  kscd-broadcasts a message $m$,
it terminates its kscd-broadcast invocation and kscd-delivers a message
set containing $m$.
\end{lemma}

\begin{proofL}
If a non-faulty process $p_i$ kscd-broadcasts a message $m$,
it adds $m$  to $\MEM$. The fact it  kscd-delivers $m$ follows
from Lemma~\ref{lemma-kSCD-no-blocking}. 

As (once computed at line~\ref{nRW-02}) $\todeliver1_i$ remains
constant until $p_i$'s next invocation of kscd-broadcast,  
and  $\delivered_i$ increases  when   $\todeliver2_i\neq\emptyset$, 
it also follows
from Lemma~\ref{lemma-kSCD-no-blocking} that the set
$\todeliver1_i\setminus \delivered_i$ decreases and becomes eventually empty.
When this occurs $p_i$ returns from its kscd-broadcast invocation. 
\renewcommand{\toto}{lemma-kSCD-termination-1} 
\end{proofL}

\begin{theorem}
\label{theorem-kSCD}
Algorithm~{\em\ref{algo:rksa-snap-to-kscd}} implements {\em KSCD-broadcast}
from $k$-set agreement and snapshot objects.  
\end{theorem} 

\begin{proofT}
The proof follows from
Lemma~\ref{lemma-kSCD-bounded}   (KSCD-Bounded property), 
Lemma~\ref{lemma-kSCD-Validity}  (KSCD-Validity property), 
Lemma~\ref{lemma:kSCD-no-duplication} (KSCD-Integrity property),
Lemma~\ref{lemma-kSCD-msg-ordering}  (KSCD-Ordering property),
Lemma~\ref{lemma-kSCD-termination-1} (KSCD-Termination-1 property), and
Lemma~\ref{lemma-kSCD-termination-2} (KSCD-Termination-2 property).
\renewcommand{\toto}{theorem-kSCD} 
\end{proofT}

\section{Conclusion}
\label{sec:conclusion}

This paper has introduced a new communication abstraction, denoted
$k$-BO-broadcast, which  matches $k$-set agreement in
asynchronous crash-prone wait-free systems. In the case $k=1$
(consensus is $1$-set agreement), $1$-BO-broadcast boils down to Total
Order broadcast.  ``Capture'' means here that (i) $k$-set agreement
can be solved in any system model providing the $k$-BO-broadcast
abstraction, and (ii) $k$-BO-broadcast can be implemented from $k$-set
agreement in any system model providing snapshot objects. It follows
that, when considering asynchronous crash-prone wait-free systems
where basic  communication\\
\begin{minipage}[l]{8.2cm}
\scalebox{0.23}{\begin{picture}(0,0)%
\includegraphics{fig-global-picture-4.pdf}%
\end{picture}%
\setlength{\unitlength}{4144sp}%
\begingroup\makeatletter\ifx\SetFigFont\undefined%
\gdef\SetFigFont#1#2#3#4#5{%
  \reset@font\fontsize{#1}{#2pt}%
  \fontfamily{#3}\fontseries{#4}\fontshape{#5}%
  \selectfont}%
\fi\endgroup%
\begin{picture}(13410,9503)(557,-10625)
\put(8641,-5551){\makebox(0,0)[lb]{\smash{{\SetFigFont{29}{34.8}{\familydefault}{\mddefault}{\updefault}{\color[rgb]{0,0,0}SCD-based Algo. in \cite{IMPR17}}%
}}}}
\put(6211,-2581){\makebox(0,0)[lb]{\smash{{\SetFigFont{34}{40.8}{\familydefault}{\mddefault}{\updefault}{\color[rgb]{0,0,0}$k$-SCD}%
}}}}
\put(3241,-5911){\makebox(0,0)[lb]{\smash{{\SetFigFont{34}{40.8}{\familydefault}{\mddefault}{\updefault}{\color[rgb]{0,0,0}$k$-BO}%
}}}}
\put(3106,-8836){\makebox(0,0)[lb]{\smash{{\SetFigFont{34}{40.8}{\familydefault}{\mddefault}{\updefault}{\color[rgb]{0,0,0}$k$-SA}%
}}}}
\put(8596,-8836){\makebox(0,0)[lb]{\smash{{\SetFigFont{34}{40.8}{\familydefault}{\mddefault}{\updefault}{\color[rgb]{0,0,0}Snapshot}%
}}}}
\put(3241,-4066){\makebox(0,0)[lb]{\smash{{\SetFigFont{29}{34.8}{\familydefault}{\mddefault}{\updefault}{\color[rgb]{0,0,0}Algo.~ \ref{algo:kscd-to-kbo}}%
}}}}
\put(11566,-3616){\makebox(0,0)[lb]{\smash{{\SetFigFont{29}{34.8}{\familydefault}{\mddefault}{\updefault}{\color[rgb]{0,0,0}Algo.~\ref{algo:rksa-snap-to-kscd}}%
}}}}
\put(1891,-7126){\makebox(0,0)[lb]{\smash{{\SetFigFont{29}{34.8}{\familydefault}{\mddefault}{\updefault}{\color[rgb]{0,0,0}Algo.~\ref{algo:kbo-to-rksa}}%
}}}}
\end{picture}%
}\\
\centerline{Figure 2: Detailing the  global view}
\end{minipage}
\begin{minipage}[r]{7.6cm}
 \vspace{+0.1cm} 
is through is a set of atomic read/write,  
or the asynchronous message-passing system
enriched with the failure detector $\Sigma$~\cite{BR10,DFG10}, 
$k$-BO-broadcast and $k$-set agreement are the two faces of
the same  coin: one is its communication-oriented face
while the other one is its agreement-oriented face.

From a technical point of view, a complete picture of the content of the paper
appears on the left. It is important to notice that the two constructions inside the dotted curve
are free from concurrent objects: each rests only on an underlying (appropriate)
communication abstraction. 
\end{minipage}

\Xomit{\begin{figure}[htb]
\centering{
\ifpdf
\scalebox{0.25}{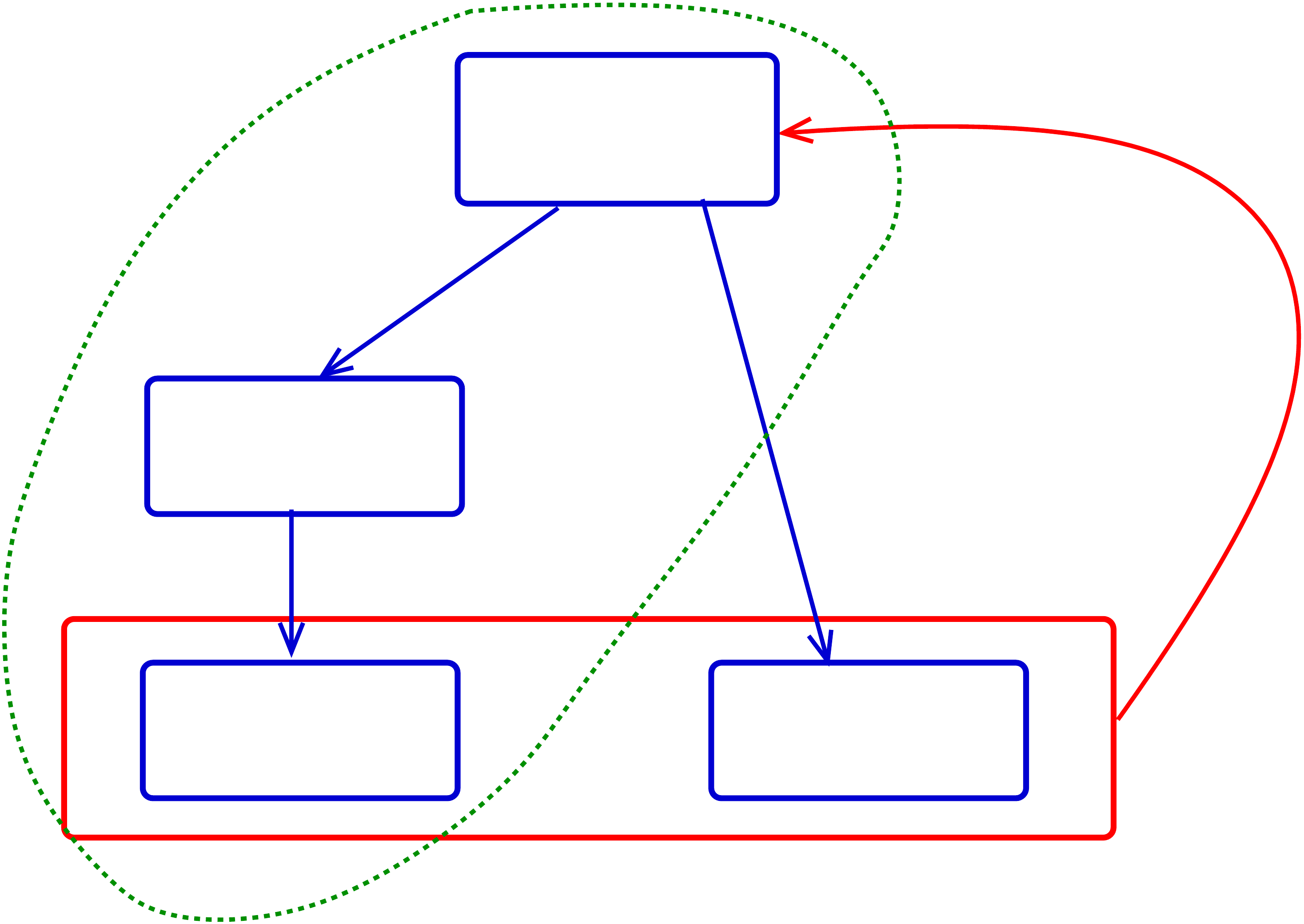}
\else
\scalebox{0.25}{\input{fig-global-picture-4.pstex_t}}
\fi
\caption{Detailing the  global view}
\label{fig:complete-picture}
}
\end{figure}

From a technical point of view, a complete picture of the content of the paper
appears in Figure~\ref{fig:complete-picture}.
It is important to notice that the two constructions inside the dotted curve
are free from concurrent objects: each rests only on an underlying (appropriate)
communication abstraction. 
}

\section*{Acknowledgments}
This work has been partially supported by the Franco-German DFG-ANR
Project 14-CE35-0010-02 DISCMAT (devoted to connections between mathematics
and distributed computing), and the French ANR project 16-CE40-0023-03
DESCARTES (devoted to layered and modular structures in distributed computing).

\newpage

\setcounter{page}{1}
\pagenumbering{roman}

\appendix

\section{Extending the scope of the result}
$k$-Simultaneous consensus ($k$-SC) was introduced in~\cite{AGRRT10}.
Each process participates in $k$ independent consensus instances, to
which it proposes the same value, until it decides in any one of them.
It is shown in~\cite{AGRRT10} that $k$-simultaneous consensus and
$k$-set agreement ($k$-SA) are equivalent in wait-free read/write
systems.  Hence, it follows that $k$-simultaneous consensus, $k$-set
agreement, $k$-BO-broadcast ($k$-BO), and $k$-SCD-broadcast ($k$-SCD)
are computationally equivalent in wait-free read/write systems.  This
provides us with a larger view of the agreement power of
$k$-BO-broadcast and $k$-SCD-broadcast, for $1\leq k <n$
(see Table~\ref{fig:object-classes} which complements Table~\ref{table1}).

\begin{table}[ht]
\begin{center}
\renewcommand{\baselinestretch}{1}
\small
\begin{tabular}{|c|c|}
\hline
$k$  &   Equivalence classes\\
\hline
\hline
$k=1$    &  $1$-B0, $1$-SCD, $1$-SA (consensus), Total order broadcast\\
\hline
$2\leq k \leq (n-2)$  & .................................. \\
\hline
$k=n-1$  &  $(n-1)$-B0, $(n-1)$-SCD, $(n-1)$-SA \\
\hline
$k\geq n$    &  $\emptyset$ \\
\hline 
\end{tabular}
\end{center}
\vspace{-0.3cm}
\caption{Equivalence classes in $n$-process wait-free read/write systems}
\label{fig:object-classes}
\end{table}

It is shown in~\cite{BT13,RS13} that $k$-simultaneous consensus is
computationally stronger than $k$-set agreement in wait-free
message-passing systems. While $k$-BO-broadcast captures repeated
$k$-set agreement in wait-free message-passing systems
(Algorithm~\ref{algo:kbo-to-rksa}), the previous observation motivates
the research of the communication abstraction which captures
$k$-simultaneous consensus in wait-free message-passing systems.

\end{document}